\documentclass[aps,prl,twocolumn,showpacs]{revtex4}

\usepackage{amsmath,amsfonts,amssymb,graphics,graphicx,epsfig,color,bbm,subfigure}

\newcommand{\be}{\begin{equation}}
\newcommand{\ee}{\end{equation}}
\newcommand{\bea}{\begin{eqnarray}}
\newcommand{\eea}{\end{eqnarray}}
\newcommand{\ket}[1]{|#1\rangle}
\newcommand{\bra}[1]{\langle#1|}

\newcommand{\eq}[1]{Eq.~(\ref{#1})}

\begin{document}

\title{Distributed quantum computation via optical fibres}
\author{Alessio Serafini,$^{1}$ Stefano Mancini,$^{2}$ and Sougato Bose$^{1}$}
\affiliation{$^{1}$ Department of Physics \& Astronomy, University College London, 
Gower Street, London WC1E 6BT, United Kingdom\\
$^{2}$ INFM and Dipartimento di Fisica, Universit\`a di Camerino, I-62032 Camerino, Italy}
\begin{abstract}
We investigate the possibility of realising effective quantum gates between 
two atoms in distant cavities coupled by an optical fibre. 
We show that highly reliable swap and entangling gates are achievable.
We exactly study the stability of these gates in presence of imperfections in coupling strengths
and interaction times and prove them to be robust. Moreover, we analyse 
the effect of spontaneous emission and losses and show that such gates are very promising in view  
of the high level of coherent control currently achievable in optical cavities.
\end{abstract}
\pacs{03.67.Lx, 03.67.Mn, 42.81.Qb}

\maketitle

The study of the possibilities allowed by coherent evolutions 
of quantum systems is central to quantum information 
science. Most notably, exploiting suitable coherent dynamics
to implement deterministic quantum gates between separate subsystems 
is a basic aim for quantum computation.
Several proposals have been suggested to engineer entanglement or
quantum communication between atoms trapped in distant optical cavities, 
either through direct linking of the cavities \cite{origin,
pellizzari97,vanenk99,parkins03}, 
or through detection of leaking photons \cite{detection1,duan03}. 
The realisation 
of quantum gates between distant qubits in quantum optical settings 
has also been recently envisaged \cite{compudet,cluster}.
Such proposals are very promising and highly inventive. 
However, they are either probabilistic 
or relying on accurately tailored sequences of pulses
(thereby requiring a considerable degree of control).
In this paper, an alternative to such schemes is proposed, 
with a particular focus on the implementation of distributed quantum computation.
To this aim, we investigate the possibility of 
realising deterministic gates between two-level atoms 
in separate optical cavities, 
through a coherent resonant coupling mediated by an optical fibre.
The only control required would be the synchronised switching on and off 
of the atom-field interactions in the distant cavities, achievable through simple control pulses.
The study of such a system (which 
would constitute the basic cell of scalable optical networks) is crucial
in view of the outstanding improvements currently achieved
in the control of single atoms trapped
in optical cavities \cite{exp} and of the recent realisation of 
microfabricated cavity-fibre systems \cite{trupke}. 

In the considered system 
the interaction between the qubits 
is mediated by the bosonic light field.
It has been showed that, in principle, an exact deterministic gate
may be realized if the interaction between two qubits is 
mediated by another two-level system through 
XY nearest neighbour interactions \cite{yung04}. 
If the central system is a bosonic field though, interacting 
with the two qubits through a rotating wave Hamiltonian, a perfect
gate is not possible, as the Rabi frequencies in the two- and single-
excitation subspaces are no longer commensurate and the mediating field 
does not exactly decouple from the qubits at short enough times.
However, as we will show,
times do exist for which the qubits are decoupled from the field 
at a high degree of accuracy. The resulting effective dynamics of the 
two qubits can then be described in terms of quantum operations which 
approximate unitary gates with a high fidelity. 
The discrepancy between such approximate 
gates and the desired unitary ones would be negligeable 
with respect
to the errors involved by an experimental implementation
of the scheme.

We consider two two-level atoms in distant optical cavities,
interacting with the local cavity fields through dipole interactions
in rotating wave approximation. 
The two cavities will be henceforth 
labeled by the indexes $1$ and $2$. We will allow for a detuning $\Delta$
of the transition of atom $2$ from the resonance frequency $\omega$ of the cavities 
(whereas atom $1$ will be assumed to be at resonance).
The cavities are connected by an optical 
fibre, whose coupling to the modes of the cavities may be modeled
by the interaction Hamiltonian 
$
H_{If} = \sum_{j=1}^{\infty}\nu_j\left[b_j(a_1^{\dag}+(-1)^j\,{\rm e}^{i\varphi}a_2^{\dag})
+\,{\rm h.c.}\right]
$ \cite{pellizzari97},
where $b_i$ are the modes of the fibre, $a_1$ and $a_2$ are the cavities' modes, 
$\nu_i$ is the coupling strength with the fibre mode $i$ and the phase $\varphi$ is due 
to the propagation of the field through the fibre of length $l$: $\varphi=2\pi\omega l/c$
\cite{haminote}. 

Now, let $\bar\nu$ be the decay rate of the cavities' fields into a
{\em continuum} of fibre modes. Taking into account a finite length $l$ of the 
fibre implies a quantization of the modes of the fibre with frequency spacing given by 
$2\pi c/l$.
One has then that the number of modes which would significantly interact with 
the cavities' modes is of the order of $n=(l\bar\nu)/(2\pi c)$ \cite{pellizzari97}. 
We will focus here on the case
$n\lesssim 1$, for which essentially only one (resonant) mode of the fibre will interact 
with the cavity modes (``short fibre limit'') \cite{filternote}. 
Notice that such a regime applies in most realistic experimental situations: for instance, 
$l\lesssim 1\,{\rm m}$ and $\bar\nu\simeq 1\,{\rm GHz}$ (natural units are adopted with $\hbar=1$)
are in the proper range. 
We recall that the coupling $\nu$ to the modes of a fibre of finite length
can be estimated as $\nu\simeq \sqrt{4\pi\bar\nu c/l}$.
Let us also notice that the coupling strength $\nu$ can be increased by decreasing the reflectivity 
of the cavity mirror connected to the fibre.
In the specified limit, the Hamiltonian $H_{If}$ reduces to $H_{f}$
\be
H_f = \nu\left[b(a_1^{\dag} + {\rm e}^{i\varphi} a_2^{\dag})+\,{\rm h.c.}\right] \; ,
\label{hamili}
\ee
where $b$ is the resonant mode of the fibre.
The total Hamiltonian of the composite system can be written, in a frame rotating at 
frequency $\omega$, as
\be
H =
\Delta \ket{1_2}\bra{1_2} + 
\sum_{j=1}^{2} (g_j \ket{0_j}\bra{1_j}a_j^{\dag}+\,{\rm h.c.}) 
+ H_f\, , \label{hami1}
\ee
where $\ket{1_j}$ and $\ket{0_j}$ are the excited and ground states 
of atom $j$, $g_j$ is the dipole 
coupling between atom and field in cavity $j$  
(generally complex, as local coupling phases, depending on the 
positions of the atoms in the cavities, might be present) and 
$\Delta$ is the detuning of the transition of atom $2$.
The addressed system is thus equivalent to two qubits 
connected by a chain of three harmonic oscillators.
For ease of notation, let us also define $g\equiv|g_1|$, 
$\delta\equiv|g_2|-|g_1|$ and $\sigma^{-}_{j}=\ket{g_j}\bra{e_j}$ for
$j=1,2$. 

Before proceeding, let us remark an interesting
feature of the Hamiltonian $H$, which unveils some significant insight about 
the dynamics we intend to study. 
Let us consider the normal modes $c$ and $c_{\mp}$ of the 
three interacting bosonic modes. One has $c=(a_1-\,{\rm e}^{-i\varphi}a_2)/\sqrt{2}$, 
with frequency $\omega$, and $c_{\mp}=(a_1+\,{\rm e}^{-i\varphi}a_2\mp\sqrt{2}b)/2$, 
with frequencies $\omega\mp\sqrt{2}\nu$. The three normal modes 
are not coupled with each other but interact with the atoms because of the  
contributions of the cavity fields. However, for $\nu\gg|g_j|$, 
the interaction of the atoms with the non resonant 
modes is highly suppressed (it is essentially limited to the second order in the Dyson series) 
and the system reduces to two qubits resonantly coupled through a single harmonic oscillator.
Remarkably, as the 
dominant interacting mode $c$ has no contribution from the fibre mode $b$, the system 
gets in this instance insensitive to fibre losses. 
On the other hand, note that
fulfilling the condition $\nu\gg|g_j|$ might require weak couplings, thus 
implying larger operating times.

Let us now discuss the computational 
possibilities allowed by the coherent evolution described by the Hamiltonian (\ref{hami1}). 
To this aim, we will be interested in the reduced dynamics 
of the two distant atoms. We will assume that the system can 
be `initialized' bringing all the field modes in the vacuum state and allowing for any initial 
state of the qubits. The Hamiltonian $H$ clearly conserves the number of global excitations and,
for our aims, one can restrict to the zero-, single- and two-excitation subspaces. 
The quantum operation describing the effective dynamics of the atoms
can thus be exactly worked out determining its Kraus operators for any values of $\nu$, $g_j$
and $\Delta$.
Denoting by 
$\ket{ijk}$ the state of the field 
given by the number state $i$ in the mode of cavity $1$, $k$ in the mode of cavity $2$ and $j$ 
in the fibre mode, one has 
$E_{ijk}(t)=\bra{ijk}\exp{(-iHt)}\ket{000}$ for $i,j,k=0,1,2$ and the state of the atoms
$\varrho (t)$ is given by 
$\varrho(t)=\sum_{i,j,k=0}^2 E_{ijk}(t)\varrho(0)E_{ijk}^{\dag}(t)$.
In particular, we are interested in singling out ``decoupling times'' 
at which the state of the atoms will be highly decoupled 
from the light field so that their evolution will be approximately unitary. 
At such times the field has a very high probability 
of being in the vacuum state in both the single- and two-excitation subspaces (the global vacuum is a 
trivial eigenvector of $H$). 
This condition is fulfilled when the Kraus operators $E_{ijk}\simeq0$ for $i,j,k\neq 0$, 
so that the Kraus operator $E_{000}$ approximates a unitary evolution.
More precisely, the fidelity of a Kraus operation $\{E_{ijk}\}$ emulating a unitary 
gate $U$ can be properly estimated as follows. 
Suppose a pure two-qubit state $\ket{\psi}$ enters the operation as input: a measure of the 
reliability of the gate is given by the overlap 
$$
f(\ket{\psi})=
\bra{\psi}U^{\dag}\left(\sum_{i,j,k=0}^{2}E_{ijk}\ket{\psi}\bra{\psi}E_{ijk}^{\dag}\right)
U\ket{\psi} \, .
$$
The fidelity $F$ of the gate may then be 
obtained by averaging over all pure input states:
$
F\equiv\langle f(\ket{\psi})
\rangle_{\ket{\psi}} 
$.

\begin{figure}[t!]
\begin{center}
\includegraphics[scale=1]{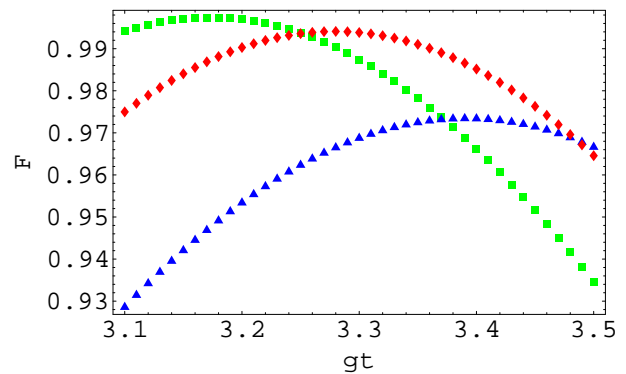}
\caption{Fidelities of an emulated swap gate as a function of time. 
The gate is obtained for $|g_1|=|g_2|=g$ and $\Delta=0$;
the diamonds refer to $\nu/g=1.1$, the squares refer to $\nu/g=1.2$, while the triangles
refer to $\nu/g=1$. 
All the quantities plotted are dimensionless.\label{swapfig}}
\end{center}
\end{figure}

Setting $\Delta=0$, $\delta=0$ and $g\simeq\nu$ yields 
a highly reliable swap gate at the decoupling time $t\simeq \pi/g$.
The fidelity of the proposed swap operation 
is shown in Fig.~\ref{swapfig}. As apparent, such a fidelity can exceed the value $0.99$
and is remarkably stable with respect to possible imperfections in the coupling 
strengths and in the temporal resolution needed to switch off the interaction once the desired 
evolution is achieved. 
Let us remark that the values $g\simeq\nu\simeq 1 \,{\rm GHz}$ 
(at hand with present technology in
optical cavities) would grant an operating time $\tau\simeq 1\,{\rm Ns}$.
We also report that, after a time $t\simeq 3.4/g$, 
a swap gate with fidelity $F\simeq 0.98$ can be obtained 
for $\nu\simeq 100g$ (and $\Delta=\delta=0$), {\em i.e.}~in the range of parameters 
for which the system gets insensitive to fibre losses. 
This agreeable advantage is thus achieved by allowing a longer operating time (due to the 
condition on $g$) and a slightly lower (but still almost perfect) fidelity.

\begin{figure}[tb!]
\begin{center}
\includegraphics[scale=1]{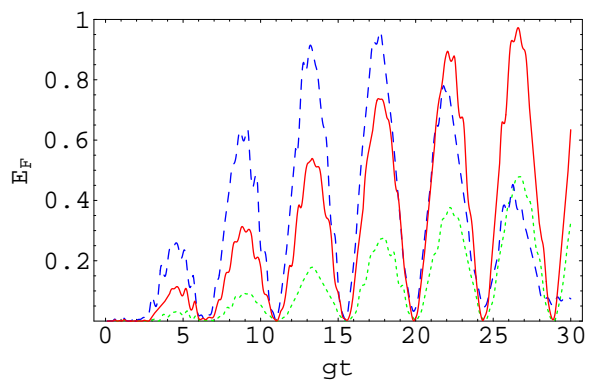}
\caption{Entanglement of formation in ebits as a function of time (in units $g^{-1}$) for 
$\nu=100g=10\Delta$ and $\delta=g$ (dashed line), $\delta=0.5g$ (continuous line) and
$\delta=0$ (dotted line). At the peaks, {\sc cphase} gates are emulated.\label{eform}}
\end{center}
\end{figure}

Moreover, this model allows for 
a reliable emulation of an entangling gate.
To fix ideas, we focus on a `controlled-phase' ({\sc cphase}) gate between the two qubits, 
described by the unitary matrix $U_{\vartheta}$ in the computational basis:
$U_{\vartheta}=\,{\rm Diag}\,(1,1,1,\,{\rm e}^{i\vartheta})$.
This gate is equivalent, up to local unitaries, to the gates 
${\rm Diag}\,(1,\,{\rm e}^{i\vartheta_1},\,{\rm e}^{i\vartheta_2},
\,{\rm e}^{i\vartheta+\vartheta_1+\vartheta_2})$ for any $\vartheta_1$,
$\vartheta_2\in [0,2\pi]$, since the phases $\vartheta_1$ and $\vartheta_{2}$
can be cancelled out by local phase gates. 
We will thus henceforth refer to all such gates as ``{\sc cphase}'' gates.
The entangling power of such gates increases 
as the phase $\vartheta$ increases between $0$ and $\pi$ (for which a controlled-$Z$ gate is achieved). 
Let us also recall that any of these 
entangling gates, together with local unitary operations, make up a universal set of gates
(as {\em any} two-mode gate can be recovered as a proper combination of the entangling gate and 
of local gates \cite{bremner02}). 
The symmetry of the Hamiltonian (crucial in realising a swap gate), must be broken here 
because it prevents a 
phase $\vartheta$ to appear at decoupling times.
In point of fact, if the transition of atom $2$ is detuned ({\em e.g.}, by Stark or Zeeman effect), 
a phase does arise, 
thus allowing for an effective entangling gate.
Reliable decouplings allowing to emulate such a gate are achieved for $\nu\gg |g_j|$,
for which the fibre is ``bypassed'' and fibre losses do not affect the performance of the gate. 
\begin{figure}[t!]
\begin{center}
\includegraphics[scale=1]{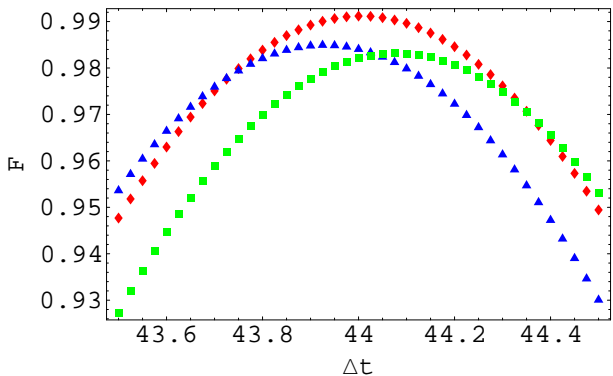}
\caption{Fidelities of an emulated {\sc cphase} gate ($U_{0.15\pi}$) as a function of time. 
The diamonds refer to $\nu/\Delta=10$, $|g_1|/\Delta=0.1$ and 
$|g_2|/\Delta=0.15$;
the squares and the triangles refer, respectively, 
to a relative variation of $-5\%$ and $+5\%$ in $|g_1|$, $|g_2|$ and $\nu$. 
The fidelities of the successive (more entangling) {\sc cphase} gates are similar. 
All the quantities plotted are dimensionless.\label{phasefig}}
\end{center}
\end{figure}
For $\nu\simeq100g\simeq200\delta\simeq 10\Delta$ a sequence of {\sc cphase} 
gates -- separated by a period of about $4.4g^{-1}$ -- with increasing 
$\vartheta$ (ranging from $\vartheta\simeq0.15\pi$ to $\vartheta\simeq0.93\pi$) 
is emulated. 
The most entangling {\sc cphase} gate ($U_{0.93\pi}$) 
is achieved after six ``Rabi-like'' oscillations 
in the two excitation subspace.
The fidelity $F$ of the emulated gate exceeds the value $0.99$.
Its stability is demonstrated in Fig.~\ref{phasefig}. 
The operating time of the gates would range, 
for $\nu\simeq 10 {\rm GHz}$, 
from $3\mu{\rm s}$ to $0.3 \mu{\rm s}$, according to the desired entangling power.
Figure \ref{eform} shows the entanglement of formation between the two atoms generated 
for an initial state $(\ket{0}+\ket{1})\otimes(\ket{0}+\ket{1})/2$ (which gets 
maximally entangled if processed by a controlled-$Z$) with several choices of parameters.
As apparent, a speed-up in the creation of entanglement is achieved by increasing the 
relative difference $\delta/g$. 
However, too large differences ($\delta/g\gtrsim 0.5$) affect the fidelity and 
stability of the emulated gate and thus, while advantageous for building up entanglement, 
are not convenient to perform actual computation.

We now take into account dissipation due to spontaneus emission of the atoms 
and to cavity and fibre losses.
The global system is then governed, in Schr\"odinger picture, by the following master equation
\be
\dot\varrho = -i[H,\varrho] + \frac{\gamma}{2} \sum_{j=1}^2 L[a_{j}]\varrho 
+ \frac{\kappa}{2} \sum_{j=1}^2 L[\sigma^{-}_{j}]\varrho +\frac{\beta}{2} L[b]\varrho \, ,  \label{master}
\ee
where the superoperator $L[\hat{o}]$ 
is defined as $L[\hat{o}]=2\hat o\varrho\hat{o}^{\dag}-\hat{o}^{\dag}\hat{o}\varrho
-\varrho\hat{o}^{\dag}\hat{o}$ for operator $\hat{o}$ and $\kappa$, $\gamma$ and $\beta$
stand, respectively, for the spontaneous emission rate and for the cavity and fibre decay rates
(assumed for simplicity to be equal in the two cavities). 
The thermal contributions of the bath have been neglected, 
as is possible at optical frequencies.
Considering decoherence analytically for one excitation and numerically for two excitations
(by integrating \eq{master}),
the operator tomography of the process encompassing decoherence 
has been reconstructed in the cases interesting
for emulating gates. 

In the regime $\nu\gg|g_j|$ the fidelities of the gates have been 
consistently found to be essentially unaffected by fibre losses. 
In general, moreover, the `direct' effect of spontaneous emission proves to be more relevant 
than the `indirect' effect of cavity losses.
For the swap gate with $\nu\simeq1.2g$ (with maximum fidelity $F\simeq0.997$ without dissipation),
the maximum fidelity drops to $F\simeq0.956$ for $\kappa=10^{-2}g$, thus allowing 
for a still relatively reliable gate, while a fidelity $F\simeq0.989$ is 
maintained for $\kappa=\gamma=\beta=10^{-3}g$. Lower decay rate leaves the 
gate virtually unaffected, while higher rates completely spoils it.
Notice that values permitting an effective swap would be already at hand for rubidium atoms 
in integrated fibre-cavity systems (see~data from Ref.~\cite{trupke}, 
with length of the cavity $L\simeq100\mu{\rm m}$).
The case $\nu=100g=200\delta=10\Delta$, selected to demonstrate the possibility 
of a {\sc cphase} gate, proved to be slightly more sensitive to spontaneous emission and cavity losses.
Let us focus on the first gate (after one Rabi-like oscillation): 
for $\kappa=10^{-2}g$, the fidelity of the gate falls to $F\simeq0.93$ 
(in which case the fidelity of the optimal most entangling gate, achieved after six oscillations, 
is completely spoiled), while for $\kappa=\gamma=10^{-3}g$ (recall that this regime 
is insensitive to fibre losses), the fidelity of the first gate is still $F\simeq0.97$. 
Generally,
decay rates as low as $10^{-4}g$ have a negligeable effect on the performances of the gates,
while decay rates of the order of $10^{-2}g$ 
would allow for remarkable experimental demonstrations of swap and entangling gates. 
In view of the quality attained in the fabrication of 
high-finesse optical cavities, the main technical issue left seems to be limiting the 
spontaneous emission rates. 
Hyperfine ground levels (with negligible `intrinsic' spontaneous emission rates) 
of effective two-level lambda systems
could thus be good candidates for the implementation of such computational schemes.
In fact, let us consider a lambda system 
(refer to Ref.~\cite{pellizzari97} for details), 
where one transition is driven by a laser of strength $h$ with detuning $d$ and the other 
is mediated by a mode of the field with resonant coupling $h$ (assumed for simplicity to be real 
and equal to the laser strenght). 
Let $\xi$ stand for the spontaneous emission rate of the excited level, 
which will be adiabatically eliminated under the condition $d\gg h$. 
Let us suppose to exploit such a two-level system for the proposed scheme. 
In our previous notation,  
one would have \cite{pellizzari97} $g\simeq d h^{2}/(d^2+\xi^2)$ 
and $\kappa \simeq \xi h^2/(d^2+\xi^2)$, 
with $g/\kappa\simeq d/\xi$:
a large enough detuning would thus allow to coherently implement the scheme with these 
effective two-level systems.

We have investigated the implementation of quantum computation
and entangling schemes  
for atoms trapped in distant cavities coupled by an optical fibre. 
Imperfections and dissipation 
have been considered showing that, in the short fibre regime, reliable gates with 
promising operating times could be at hand with present technology. 
Let us also mention that, in the considered system, not only entangling 
and swap gates, but also perfect quantum state transfer is possible. 
Besides, the proposed setup would also allow for the unitary generation
of cluster states between distributed atoms or ions \cite{cluster}, 
and could thus find application not only in gate-based but
also in ``one-way'' quantum computation.
More generally,
our results strongly emphasize 
the potentialities of quantum optical systems
towards the realisation of effective quantum networking schemes.

We thank K.~Jacobs, M.~Trupke, C.~Foot, J.~Metz, Y.L.~Lim, W.~Lange and S.~Bergamini 
for useful discussions.
This research is part of QIP IRC www.qipirc.org (GR/S82176/01).
SM thanks UCL Department of Physics \& Astronomy for 
hospitality (funded through the grant GR/S62796/01).


\begin{thebibliography}{99}
\bibitem{haminote}{Actually, the description of the system in terms of distinct
cavity and fibre modes -- interacting through the Hamiltonian $H_{If}$ -- 
holds for high finesse cavities and near resonant operations over time scales 
long with respect to the fibre's round-trip time (see Ref.~\cite{vanenk99}). 
All these assumptions are matched by the system at hand in the operating conditions 
we will consider 
(in particular in the ``short fibre limit'', see next paragraph)}.
\bibitem{filternote} Let us remark that the fibre could be also filtered in frequency to 
ensure the selection of a single interacting mode (suggestion by M.~Trupke).
\end{thebibliography}
\end{document}